# Extreme breakdown of the Einstein relation in liquid water under centrifugation

Joseph F. Wild[1], Yihan Li[1], Keyue Liang[1], Aishwarya S. Gujarathi[1], Heng Chen[2], Alex N. Halliday[2], Stephen E. Cox[2†], Yuan Yang[1†]

**Affiliations:**

[1]Department of Applied Physics and Applied Mathematics, Columbia University, New York, NY 10027, United States

[2]Lamont-Doherty Earth Observatory, Columbia University, Palisades, NY 10964, United States

[†]Corresponding authors. Email: yy2664@columbia.edu, sec2125@columbia.edu

## Abstract

We present evidence that the Einstein relation (ER) breaks down completely in pure water and dilute aqueous solutions under strong centrifugation fields at 40 °C. Isotopologues (e.g., $H_2^{18}O$) and solutes migrate at a speed of only ~5% of that predicted based on the ER. The ER is restored with the addition of solutes above a transition concentration ($c_t$). We further discovered a new scaling law between the solute's partial molar density, the centrifugal acceleration, and $c_t$, which can be quantitatively described by a two-phase model in analog to the Avrami model for phase transformation. The breakdown may stem from long-range dipole interactions or the hydrogen bond network in water, which are disrupted by the presence of solutes. This report shows that studying transport under centrifugation can be a new strategy to understand fundamental transport properties and complex interactions in liquids.

1. **Introduction**

Liquid water is arguably the most important chemical in the universe as it sustains life (1, 2). It is chemically simple but possesses unusual physical properties, such as anomalous expansion (3, 4), the Jones-Ray effect (5, 6), long-range orientational ordering (7-11), and phonon-like propagation of molecular vibration (12). These anomalies originate from water's unique structures, such as the imperfect tetrahedral angles, hydrogen bonds, and strong dipole moment (3). Fundamental understanding of these unusual properties is critical to the countless areas of technology and biology that are reliant on water.

One intriguing and puzzling phenomenon in water is the breakdown of the Stokes-Einstein Relation (SER) in supercooled water (13-15). SER states that $D\eta/T$ is a constant, where $D$ is the diffusion coefficient of a dissolved species $s$ in water, $\eta$ is the viscosity of water, and $T$ is the temperature (16). This results from the more general Einstein Relation (ER) in the form of $D/u = RT$, where $R$ is the ideal gas constant, and $u$ is the mobility of the species $s$ under an external field (e.g., electrical, centrifugal) (17). SER is a special case of ER with the assumption that the Stokes' law holds (that the frictional force in a liquid is proportional to viscosity). In water, $D\eta/T$ is found to be a constant as predicted at $T > 50\ °C$, but it gradually increases with decreasing temperature. The deviation is <5% between 20 and 50 °C, ~12% at 0 °C, and ~60% at -30 °C (Fig. S1) (18). This breakdown is hypothesized to arise from a liquid-liquid phase transition in water (13, 14, 19).

Here, we report a strikingly more substantial breakdown of ER in water at 40°C in a centrifugal field, where the centrifugal mobility of $H_2^{18}O$ tracers decreases such that $D/u$ is ~ $20 \times RT$ in pure water, indicating a deviation of ~1900% from ER (Fig. 1A). This breakdown disappears with the addition of various solutes above a critical concentration ($c_t$). Intriguingly, we find that $c_t$ is proportional to the square of the partial molar density of the solute ($\partial \rho_{soln}/\partial c_i$ where $\rho_{soln}$ is the solution density and $c_i$ is the concentration of solute $i$), such as 0.9 mM for CsI, 3.0 mM for $MgSO_4$, 350 mM for EtOH, and 6.0 M for $H_2^{18}O$. Moreover, we find that this ER breakdown applies to all chemical species in water - not only isotopes in $H_2O$, but also the solutes themselves.

Such a scaling law is unknown to the best of our knowledge. The partial molar density dependence suggests that the ER breakdown may stem from molecular vibration (e.g., phonon-like hydrogen-bond network (12) or long-range dipole interactions (20)). Our results unveil a new exotic behavior of water, which suggests an unknown structure in water. Moreover, this report shows that studying transport under centrifugation can create a new paradigm to understand complex interactions in liquids.

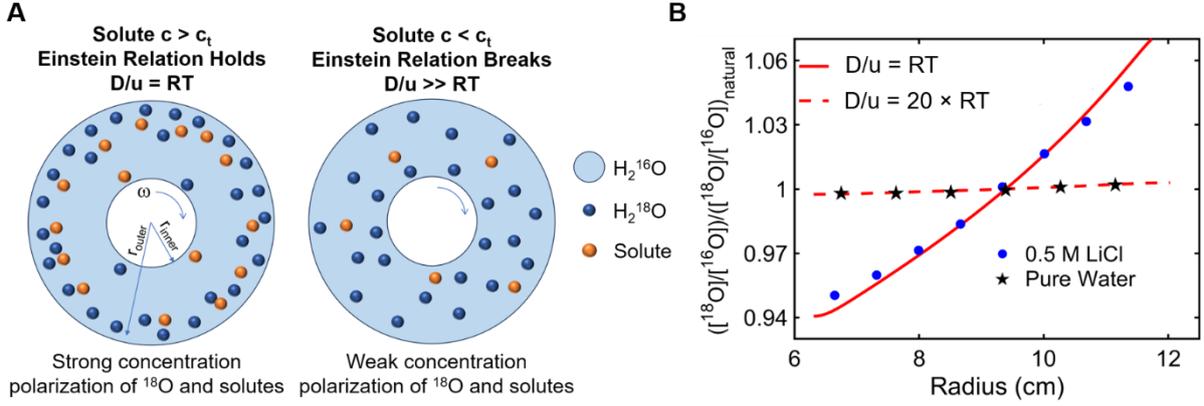

**Figure 1** – (A) Left: The Einstein relation leads to substantial separation of isotopes and solutes in high-speed centrifugation, which is observed in aqueous solutions with solutes above a transition concentration ($c_t$). Right: Breakdown of the Einstein relation results in much less separation of isotopes and solutes in centrifugation, which is observed in pure water and water with aqueous solutions with solutes below $c_t$. The breakdown is suspected to result from an unclear physical origin related to long-range correlations or molecular vibration in water. (B) The normalized spatial concentration distribution of [$^{18}$O]/[$^{16}$O] in pure water (black stars) and in 0.5 M LiCl aqueous solution (blue dots) after 48 hours of centrifugation at 60 kRPM. The solid line shows the predicted relationship, $D/u = RT$, which closely matches only the 0.5 M LiCl experiment. [$^{18}$O]/[$^{16}$O] is normalized to the ratio before centrifugation, which is denoted as ([$^{18}$O]/[$^{16}$O])$_{natural}$.

## 2. Results

### 2.1 Centrifugation modeling

Centrifugation provides a universal method to obtain field-driven mobility no matter whether the target species is neutral or charged. When a tracer (e.g., H$_2^{18}$O in H$_2^{16}$O) is subjected to a centrifugal field in a dilute solution, its transport satisfies

$$\vec{J} = -D\vec{\nabla}c + uc\omega^2\vec{r}\,M(1 - \bar{v}\rho_{solvent}) = -D\vec{\nabla}c + uc\omega^2\vec{r}\frac{\partial \rho_{soln}}{\partial c} \quad (1)$$

$$\frac{\partial c}{\partial t} = -\nabla \cdot \vec{J} \quad (2)$$

Where $\vec{J}$, $D$, $c$, $u$, $M$ are the molar flux, diffusion coefficient, molar concentration, centrifugal mobility, and molar mass of the tracer, respectively. $\bar{v}$ is the partial specific volume of the tracer. $\omega$ is the rotational velocity, $\vec{r}$ is the radial vector from the rotation axis, $\rho_{solvent}$ and $\rho_{\text{soln}}$ are the solvent and solution density at $\vec{r}$, respectively, and $t$ is time (Section S3 for details). If the ER holds, $D/u = RT$.

In this paper, we use $^{18}$O as a neutral tracer to study migration under a centrifugal field. The separation factor of $^{18}$O/$^{16}$O at equilibrium $\alpha_o$, defined in Eq. 3 by solving $\vec{J} = 0$, gives a direct measure of $D/u$ to determine whether the ER holds (21).

$$\alpha_o = \frac{\left(c_{^{18}O}/c_{^{16}O}\right)_{\text{outer,eq}}}{\left(c_{^{18}O}/c_{^{16}O}\right)_{\text{inner,eq}}} = \exp\left(\frac{u}{2D} \times \omega^2(M_{^{18}O} - M_{^{16}O})(r_{\text{outer}}^2 - r_{\text{inner}}^2)\right) \quad (3)$$

The subscripts *inner* and *outer* represent the inner and outer radii of the centrifuge tube, respectively (Section S3 for details). The transient separation factor $\alpha$, defined in the same way as $\alpha_o$ except at a transient time, also obeys $\ln(\alpha) \propto (D/u)^{-1}$, so comparing experimental results after a fixed centrifugation time equally reveals the variation of $D/u$ (Section S5 for details) (22-24).

**2.2 Experimental results on Einstein Relation breakdown**

In a typical experiment, we centrifuge 18.2 MΩ deionized water (pure water) at 60 kRPM at 40 °C for various times with a lab-scale ultracentrifuge. Then, we measure the isotope ratio $^{18}$O/$^{16}$O at $r_{\text{inner}}$ and $r_{\text{outer}}$ to obtain the transient isotope separation factor, $\alpha$ (Section S1 for details). If the ER holds, simulation shows that we should observe $\alpha$ of 1.090 after 24 hr and 1.129 after 48 hr (solid red line in Fig. 1B), which is consistent with experimental results in 0.5 M LiCl aqueous solution (e.g., blue dots in Fig. 1B). The ER is also validated in centrifuging various other salt solution in our past publication (see Section S5 for details)(24).

Instead, we obtain α of only 1.0040±0.0012 and 1.0062±0.0013 after 24 h and 48 h, respectively, in pure water (black stars in Fig. 1B). This corresponds to $D/u$ of (21.7±6.6) × $RT$ and (19.6±4.1) × $RT$, respectively, representing a factor of ~20 deviation from the ER. Additionally, the internal distribution of isotopes throughout the centrifuge tube was found to be consistent with $D$ in literature (3.2 × 10$^{-9}$ m$^2$/s at 40°C) (25) and a centrifugal mobility, $u$, reduced by a factor of ~20, which is shown as the red dashed line in Fig. 1B. These deviations are 1-2 orders of magnitude higher than those previously observed in the breakdown of SER in supercooled water (14, 26, 27).

Such a large deviation is puzzling since all previous studies shows that the SER breakdown in water is <5% above 20°C (14, 18). More puzzling is that the transport and thermodynamic properties of very dilute water have been thoroughly studied and are self-consistent around ambient temperatures - for example, the self-diffusion coefficient, ionic mobility, activity coefficient, and heat capacity (28-32).

To better understand this phenomenon, we investigated the effect of solutes. We tested 12 different solutes spanning salts, neutral molecules, and isotopes. As shown in Fig. 2A, we found that the ER breakdown phenomenon completely disappears when enough solute is added. However, the transition concentration ($c_t$), which is defined as the concentration at which $\alpha$ reaches 50% of the theoretical value, varies across four orders of magnitude, and $c_t$ is lower in general when the solute is denser. For example, $c_t$ is 0.9 mM for CsI, 3 mM for $MgSO_4$, 130 mM for LiCl, 350 mM for ethanol, and even 6.0 M for just $^{18}O$ isotopes (water with ~10.9 at% $H_2^{18}O$ and ~89.1% $H_2^{16}O$).

We further found a scaling law between $c_t$ and partial molar density of the added solute $i$ ($\partial \rho_{\text{soln}}/\partial c_i$). As shown in Fig. 2B and Table S2, most solutes, charged or neutral, lie on the line representing $c_t \sim (\partial \rho_{\text{soln}}/\partial c_i)^{-2}$. This strongly suggests that the ER breakdown and its disruption do not stem from the charge of solutes, but likely their physical mass, such as from disrupting vibrational modes or the generation of internal stresses under centrifugation. A noticeable outlier is TEGDME, which is slightly above the line. We suspect that this is due to the shape of the molecule being more linear and far from spherical, which causes anisotropic vibrational effects. To the best of our knowledge, such a scaling law has not been previously reported.

Besides partial molar density, $c_t$ also shows a linear relation with the centrifugal acceleration ($g$), which is calculated as $\omega^2 r_{\text{average}}$, in the range of 0.40 – 3.6 × $10^6$ m $s^{-2}$ (Fig. 2C). 3.6 × $10^6$ m $s^{-2}$ is achieved at 60 kRPM, the upper limit of the ultracentrifuge used. On the other hand, if the rotation speed is much less than 20 kRPM (0.4 × $10^6$ m $s^{-2}$), the absolute isotope separation is greatly reduced, which is difficult to determine precisely. The results in Fig. 2C indicate that $c_t \sim g^{-1}$.

In addition to $^{18}O$, we found that the ER breakdown also applies to the solute. For example, in Fig. 2D, the solute separation factor, defined as the outer solute concentration over the inner solute concentration after centrifugation, shows the same anomaly at a low concentration. Again, the anomaly disappears and the ER restores at higher solute concentrations, mirroring the effect seen in $^{18}O$. Moreover, a solute with a larger partial molar density also gives a lower transition concentration in correspondence with the $^{18}O$ results.

These results indicate that the concentration-dependent ER breakdown is a universal phenomenon which affects the mobility of everything within the aqueous solution. More details of measurements on solute concentration can be found in Section S7.

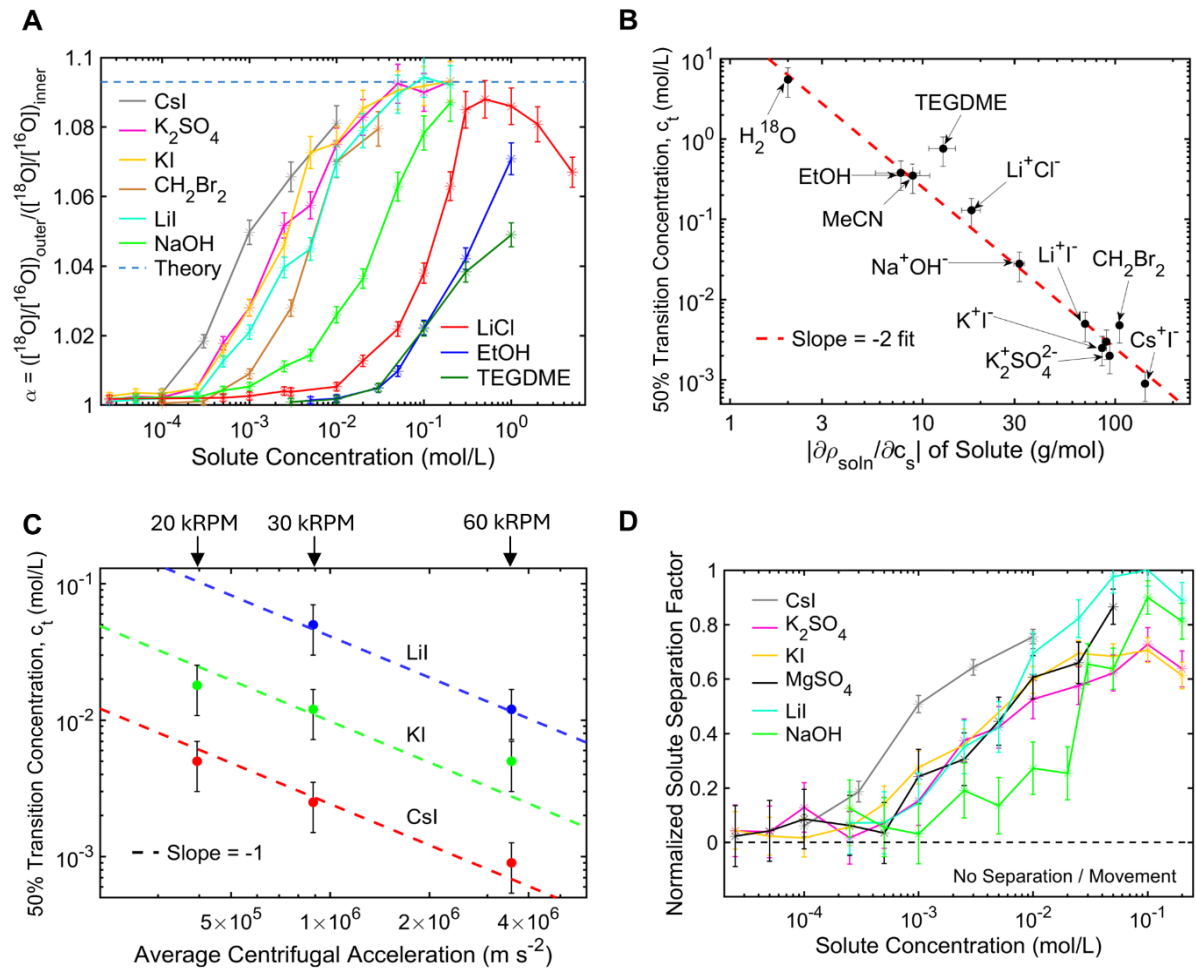

**Figure 2** - Experimental centrifugation results. (A) The separation factor α of $^{18}O/^{16}O$, which is defined as the ratio of $[^{18}O]/[^{16}O]$ at $r_{outer}$ to $[^{18}O]/[^{16}O]$ at $r_{inner}$. The top blue dashed line represents the theoretical value if the ER holds. Results for more solutes and individual results can be found in Section S6. (B) The dependence of the transition concentration ($c_t$) on the partial molar density $\partial \rho_{soln}/\partial c_i$. The dashed red line represents $c_t \sim (\partial \rho_{soln}/\partial c_i)^{-2}$. (C) The dependence of $c_t$ on the centrifugal acceleration ($\omega^2 r_{average}$), where $r_{average} = (r_{outer} + r_{inner})/2$, which is 9 cm for the centrifuge used. (D) The dependence of the solute separation factor on solute concentration in the aqueous solution. This factor is defined as the ratio of solute concentration at $r_{outer}$ to $r_{inner}$, normalized to the theoretical value if the ER is valid. All measurements were done after centrifuging for 24 hours.

### 2.3 Two-phase phenomenological model

The strongly mass-dependent $c_t$ and the charge-irrelevant behavior in Fig. 2B strongly suggest that the anomaly comes from molecular vibration in water, such as the hydrogen bond network or long-range dipolar correlation. We find that the experimental anomaly in Fig. 2 can be quantitatively described by a two-phase phenomenological model. This model assumes that an aqueous solution, with a total volume $V_{\text{total}}$, consists of two phases (Fig. 3A). The phase I, with volume $V_1$, comprises spheres around solute centers (molecules, ions, or isotopes) where the ER is valid ($D/u_1 = RT$). In this phase, the volume of spheres per mole for a specific solute $i$ ($V_{0,i}$) is

$$V_{0,i} = A_0 \left(\frac{\partial \rho_{\text{soln}}}{\partial c_i}\right)^2 g \quad (4)$$

where $A_0$ is the single fitting parameter with a value of $5.57 \times 10^{-6}$ m²·s²·mol/kg². In this model, different ions are treated as separate solute centers, and Eq. 4 indicates that the volume of a single sphere for $Cs^+$, $I^-$, $Li^+$, and EtOH is 540 nm³, 310 nm³, 5.5 nm³, 3.3 nm³, respectively, at $3.6 \times 10^6$ m s$^{-2}$ (60 kRPM in our experiments).

The phase II is mono-isotopic pure water outside of phase I, and its volume $V_2 = V_{\text{total}} - V_1$. The centrifugal mobility in phase II ($u_2$) is 0. Then, the effective centrifugal mobility of a species, $s$, in the whole solution is the volume-weighted average of these two phases.

$$u_s = \frac{u_1 V_1 + u_2 V_2}{V_1 + V_2} = \frac{D}{RT} \frac{V_1}{V_{\text{total}}} \quad (5)$$

It should be noted that $V_1$ is not simply the sum of volumes of all spheres for all solutes ($V_{\text{total}} \sum_i V_{0,i} c_i$) since the spheres may overlap. Taking the overlapped region into account, the volume fraction of phase I is $V_1/V_{\text{total}} = 1 - \exp(-\sum_i V_{0,i} c_i)$, which is analogous to the Avrami model of solid-solid phase transformations (Fig. 3A and Section S9). Therefore, the effective centrifugal mobility of species $s$ is

$$u_s = \frac{D}{RT} \frac{V_1}{V_{\text{total}}} = \frac{D}{RT} \left[1 - \exp\left(-\sum_i V_{0,i} c_i\right)\right] \quad (6)$$

Then, the transition concentration $c_t$ in an aqueous solution can be expressed as $\ln(2)/\sum_i V_{0,i}$. This single-parameter model fits experimental results in Fig. 2 very well, as shown in Fig. 3B. The corresponding values of $c_t$ match experimental values well across four orders of magnitude, and $g$ is close to experimental values over a factor of 9 (Section S8). In Fig. 3B, the deviation in LiCl at high concentrations may arise from the reduced diffusion coefficient due to high viscosity. Such success strongly indicates that the breakdown of the ER in centrifugal fields arises from mass effects such as vibrational disruption.

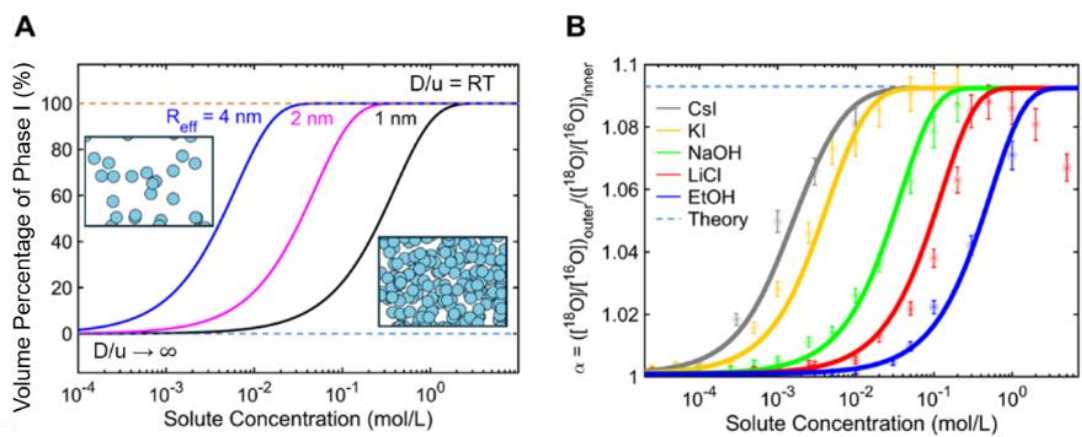

**Figure 3** – Phenomenological analysis of the experimental results using an analogy to the Avrami equation. (A) Dependence of the volume percentage of phase I (= $V_I/V_{total}$, blue region) on solute concentration with different effective radii of spheres around a solute center ($R_{eff}$). Due to overlapping between spheres, the volume of phase I shows an S-shape dependence on solute concentration. Inset: simulated 2D distribution of spheres when phase I volume is 20% (top left) and 80% (bottom right) of the total volume. (B) Experimental results of the concentration-dependent $\alpha$ in Fig. 2B, together with fitting curves from the single-fitting-parameter model described by Eq. 6. Five solutes are presented here for simplicity. Fitting results for other solutes in Fig. 2A-C are presented in Fig. S11.

### 3. Discussion

Such a breakdown of the ER under a centrifugal field has not been reported before to the best of our knowledge. While the origin is unclear, the phenomenon's dependence on partial molar density makes us suspect that it may be related to the following two mechanisms:

1) The hydrogen bond network in $H_2O$, where heavier solutes more easily disrupt the network vibrations, and thus a lower concentration of solutes is needed to break the network and restore the ER. Molecular dynamics (MD) simulations have observed optical phonon-like propagating modes in water extending 2-3 nm (12), which may act as an additional energy barrier for solute motion under an external field.

2) Long-range dipole correlations of the water molecules, which inhibit the free motion of solutes, and heavier solutes impart a greater internal stress on the water structure and thus break up the correlations. This hypothesis is supported by the repeatedly observed molecular correlations extending beyond 100 nm in liquid water (8, 11, 20). These correlations may result from rotation-translation coupling in acoustic phonons in the liquid, which would similarly act as an energy barrier for centrifugal drift and suppress free movement (10).

We also do not suspect that the observed phenomena originate from the high pressure inside water during centrifugation at 60 kRPM, which peaks at ~200 MPa. This is because past measurements show that the water diffusion coefficient only varies about 10% under pressures up to 200 MPa (28, 33), and there is no known phase transformation in liquid water below 1 GPa at 40 °C (13). Moreover, we still observe a substantial reduction of mobility when the centrifugal speed is as low as 10 kRPM (Fig. S12), which corresponds to a maximum pressure in the centrifuge tube of only 6 MPa, ~1/36 of that at 60 kRPM in Fig. 2. Therefore, the phenomenon was present regardless of pressure.

In summary, we find a significant breakdown of the Einstein relation in pure water and dilute aqueous solutions. The critical concentration for the breakdown shows a power law dependence on the partial molar density and centrifugal acceleration. Solute concentrations as low as 0.9 mM for CsI influence solution mobility by a factor of ~10 at high centrifugal accelerations. These findings suggest unknown interactions within the water structure that specifically hinder centrifugal mobility, potentially extending to other dipolar liquids. Understanding these interactions could significantly advance our knowledge of water, the liquid phase, and separation technologies.

**Funding:** This work was support in part by the United States Department of Energy, Grant Number DE-SC0022256.

**Supplementary Information:**

Materials and Methods

Sections S1 to S11

Supplementary Text

Figs. S1 to S14

Tables S1 to S5

Supplementary Materials for

**Title:** Extreme breakdown of the Einstein relation in liquid water under centrifugation

**Authors:** Joseph F. Wild[1], Yihan Li[1], Keyue Liang[1], Aishwarya S. Gujarathi[1], Heng Chen[2], Alex N. Halliday[2], Stephen E. Cox[2†], Yuan Yang[1†]

[1]Department of Applied Physics and Applied Mathematics, Columbia University, New York, NY 10027, United States

[2]Lamont-Doherty Earth Observatory, Columbia University, Palisades, NY 10964, United States

†Corresponding authors. Email: yy2664@columbia.edu, sec2125@columbia.edu

This file includes

    Materials and Methods

    Sections S1 to S11

    Supplementary Text

    Figs. S1 to S14

    Tables S1 to S5

**Section S1. Materials and Methods**

**1.1. Preparing Solutions:** All chemicals used are listed in Table S1. All solutions to be centrifuged were prepared from the same source of deionized pure water, with a temperature corrected conductivity of 0.055 µS/cm (Direct-Q 3 UV water purification system). Each centrifuge tube has a volume of around 4.0 mL. The centrifuge rotor has 6 buckets, so that up to 6 samples could run simultaneously.

**1.2. Centrifugation:** A SW 60 Ti rotor in a Beckman Optima XPN-100 Ultracentrifuge was used for all experiments. 60,000 RPM was used unless explicitly stated otherwise. The inner and outer radii are 63.1 mm and 120.3 mm, respectively. The centrifuge accelerated or decelerated at a rate of ~15,000 RPM/min. The rotor was generally initially at 15-20°C upon starting centrifugation and the heating rate was found to be around 0.4 - 0.5°C min$^{-1}$, so it would take around 1 hour to reach 40°C. At the end of the run, the temperature was set to 25°C for 1 hour at the same speed to bring the solution closer to ambient conditions and minimize convection-induced remixing upon collection. Open-top thinwall polypropylene tubes were used in all experiments.

**1.3. Sample Collection:** 0.5 mm sterile needles were used to collect the samples from the top and bottom of the centrifuge tubes immediately after the end of the run, which correspond to the inner and outer radii, respectively. This process would take around 10 minutes for all six tubes. Generally, 150-250 mg of the sample was collected in each case. The top liquid could be accessed at the top of the centrifuge tube, while the bottom liquid was accessed by carefully removing the thinwall tubes from the bucket and then slowly piercing the bottom of the tube in a twisting motion. The internal isotope distribution, as probed in Fig. 1B of the main text, involved rapidly freezing the centrifuged solution by lowering the tubes into liquid nitrogen. This quick freezing preserved the internal solute distribution. The ice was then cut into eight equal pieces along the tube's length to probe the spatial distribution of the isotopes.

**1.4. Isotopic and Concentration Measurements:** A Picarro L2130i was used for the water $^{16}O/^{18}O$ isotope measurements. This measures $H_2^{18}O$ concentrations using the isotope-dependent infrared absorption of water molecules around 7199.96 cm$^{-1}$. This method is therefore not affected by the isotopic composition of H or O atoms which may exist in the solutes, such as in TEGDME, $SO_4^{2-}$, or MeCN, rather than in water molecules itself. The instrument is designed to routinely measure seawater $H_2O$ isotopes to <1 ‰ where the salt concentrations exceed 500 mM. A Picarro-provided salt liner was used to protect the vaporizer from salt accumulation and was periodically cleaned. All samples and water standards were analyzed using the salt liners. Salt concentrations were measured using a LAQUAtwin Compact Water Quality Meter with a specified accuracy of: ±5 μS/cm (0 to 199 μS/cm), ±0.05 mS/cm (0.20 to 1.99 mS/cm), ±5 mS/cm (20 to 199 mS/cm).

### 1.5. Materials Used:

**Table S1**

| Chemical | Source | Notes |
|---|---|---|
| tetraethylene glycol dimethyl ether, ≥99% | Sigma, 172405 | Lot: BCC66172 |
| acetonitrile anhydrous, 99.8% | Sigma, 271004 | Lot: SHBL7595 |
| ethanol, 200 proof | Decon Labs, Inc, 2701 | Lot: 1922418 |
| lithium chloride, anhydrous, 99% | Sigma, 793620 | Lot: 1003676968, Stored in Ar glovebox |
| sodium hydroxide, ≥98% | Sigma, 71690 | Lot: SLBQ9677V |
| lithium iodide, anhydrous | Sigma, 818287 | Lot: S8153887 139, Stored in Ar glovebox |
| dibromomethane, 99% | Sigma, D41686 | Lot: MKCQ5255 |
| magnesium sulfate heptahydrate, ≥99.0% | Sigma, M1880 | Lot: SLBP9435V |
| potassium iodide, 99% | Sigma, 207969 | Lot: MKCH8712 |
| potassium sulfate, ≥99.0% | Sigma, P0772 | Lot: SLBP1025V |
| cesium iodide, 99.9% trace metals basis | Sigma, 202134 | Lot: 0000384169 |

**Section S2. Literature Test of the Stokes–Einstein and Stokes–Einstein–Debye Relations**

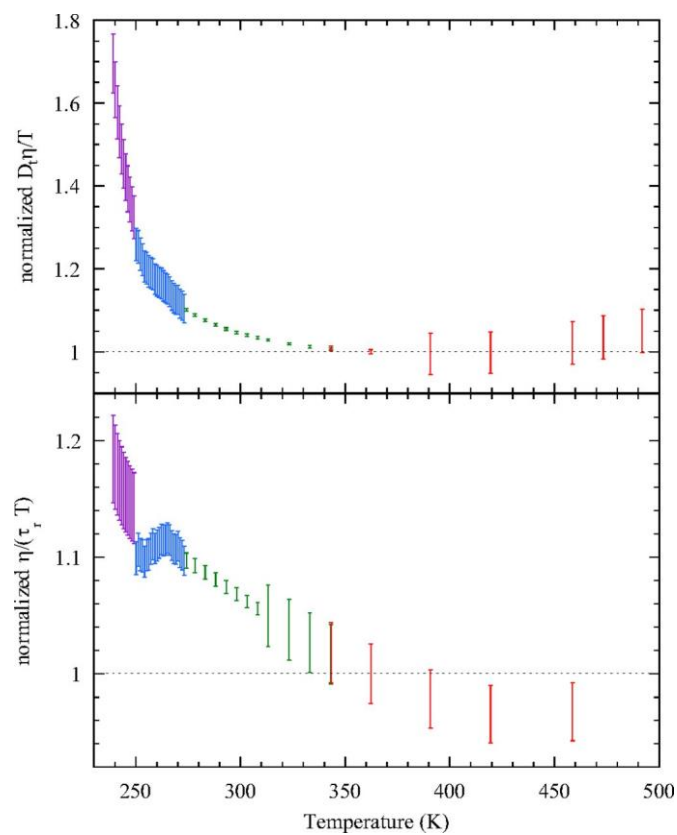

**Figure S1** - Test of the Stokes–Einstein (SE) and Stokes–Einstein–Debye (SED) relations. $D_t\eta/T$ (Upper) and $\eta/(\tau_r T)$ (Lower) are plotted as a function of temperature. $D_t$ and $\tau_r$ were calculated at the temperatures of the viscosity data using the power law fits. Only the combined uncertainty (1 SD) without the data symbol is displayed for clarity. The data were further normalized by their value at 362.25 K. SE and SED relations would thus correspond to the horizontal dotted lines. SE and SED largely hold at high temperature (300 K) with just 7% and 5% deviations, respectively, but they are violated by around 70% and 18% at low temperature (250 K), respectively. Figure reproduced from Dehaoui, A., Issenmann, B. & Caupin, F. *Viscosity of deeply supercooled water and its coupling to molecular diffusion.* PNAS 112 (2015)[1].

**Section S3. Solute Properties and Transition Concentration**

The total volume of a solution ($V_{total}$) can be written as $V_{total} = V_{solvent} + m_s \bar{v}_s$, where $V_{solvent}$ is the volume of the solvent, $m_s$ is the total mass of the solute $s$, and $\bar{v}_s$ is the partial specific volume, which is defined as $\frac{\partial V_{total}}{\partial m_s}$. Here we assume that $\bar{v}_s$ is a constant, which is a reasonable approximate for dilute solutions. The solute $s$ can be a tracer (e.g., H$_2{}^{18}$O), a dissolved neutral molecule, or a salt. Therefore:

$$V_{solvent} = V_{total} - m_s \bar{v}_s$$

Then multiplying both sides by the solvent density ($\rho_{solvent}$) gives:

$$m_{solvent} = (V_{total} - m_s \bar{v}_s)\rho_{solvent} = (V_{total} - M_s c_s V_{total} \bar{v}_s)\rho_{solvent}$$
$$= (1 - M_s c_s \bar{v}_s)V_{total}\rho_{solvent}$$

where $m_{solvent}$ is the solvent mass, $M_s$ is the molar mass of $s$, and $c_s$ is the solute concentration (mol/L). Then, the density of the solution ($\rho_{soln}$) is given by:

$$\rho_{soln} = \frac{m_s + m_{solvent}}{V_{total}} = \frac{M_s c_s V_{total} + (1 - M_s c_s \bar{v}_s)V_{total}\rho_{solvent}}{V_{total}}$$
$$= M_s c_s + (1 - M_s c_s \bar{v}_s)\rho_{solvent}$$

Therefore, the partial molar density with respect to $s$, is:

$$\frac{\partial \rho_{soln}}{\partial c_s} = M_s(1 - \bar{v}_s \rho_{solvent})$$

For two species which are chemically identical isotopes, for example H$_2{}^{16}$O and H$_2{}^{18}$O, $M_{{}^{18}O}\overline{v_{{}^{18}O}} = M_{{}^{16}O}\overline{v_{{}^{16}O}}$ since they take up identical volumes. Then the difference in partial molar density of the species is simply the difference in the molecular masses:

$$\frac{\partial \rho_{soln}}{\partial c_{{}^{18}O}} - \frac{\partial \rho_{soln}}{\partial c_{{}^{16}O}} = M_{{}^{18}O}\left(1 - \overline{v_{{}^{18}O}}\rho_{solvent}\right) - M_{{}^{16}O}\left(1 - \overline{v_{{}^{16}O}}\rho_{solvent}\right) = M_{{}^{18}O} - M_{{}^{16}O}$$

Finally, solving Eq. 1 and Eq. 2 of the main text for the equilibrium case where $\vec{J} = 0$ for all species results in the equilibrium separation factor of Eq. 3. A derivation of this result is shown below, which originates from the Supporting Information of *Wild, et al. (2025)*[2].

In equilibrium, the flux equation for two isotopic species, 1 and 2, can be written as Eqs. S1 and S2:

$$J_1 = 0 = -D_1 \frac{\partial c_1}{\partial r} + u_1 c_1 \omega^2 r M_1 (1 - \overline{v_1}\rho_{soln}) \qquad (S1)$$

$$J_2 = 0 = -D_2 \frac{\partial c_2}{\partial r} + u_2 c_2 \omega^2 r M_2 (1 - \overline{v_2}\rho_{soln}) \qquad (S2)$$

where $J_i$ is the molar flux of species $i$. $D$ is the diffusivity, $c$ is the molar concentration, and $r$ is the radius from the rotation axis. $u$ is the mobility and $\omega$ is the rotational velocity.

Since the species are isotopes of each other, $D_1 = D_2 = D$, $u_1 = u_2 = u$, and $M_1 \overline{v_1} = M_2 \overline{v_2}$. Then, multiplying Eq. S1 by $c_2$ and subtracting Eq. S2 multiplied by $c_1$ yields:

$$(S1) * c_2 - (S2) * c_1 : 0 = -D \left[ \frac{\partial c_1}{\partial r} c_2 - \frac{\partial c_2}{\partial r} c_1 \right] + u\omega^2 r c_1 c_2 (M_1 - M_2)$$

Therefore, dividing through by $c_1 c_2$:

$$D \frac{\partial c_1}{\partial r} \frac{1}{c_1} - D \frac{\partial c_2}{\partial r} \frac{1}{c_2} = D \frac{\partial \ln(c_1)}{\partial r} - D \frac{\partial \ln(c_2)}{\partial r} = u\omega^2 r (M_1 - M_2)$$

$$\rightarrow \quad \frac{\partial \ln(c_1) - \partial \ln(c_2)}{\partial r} = \frac{\partial \ln\left(\frac{c_1}{c_2}\right)}{\partial r} = \frac{u\omega^2 r (M_1 - M_2)}{D}$$

$$\rightarrow \quad \int_{r_{inner}}^{r_{outer}} d\ln\left(\frac{c_1}{c_2}\right) = \int_{r_{inner}}^{r_{outer}} \frac{u\omega^2 r (M_1 - M_2)}{D} dr$$

$$\alpha_0 = \frac{c_1(r_{outer})/c_2(r_{outer})}{c_1(r_{inner})/c_2(r_{inner})} = \exp\left(\frac{u}{2D} \times \omega^2 (M_1 - M_2)(r_{outer}^2 - r_{inner}^2)\right)$$

Table S2 gives the properties of solutes centrifuged in this study. The values for partial specific volumes are given for the dilute limit, which is justified given the concentrations generally used in this study.

**Table S2** – Properties of solutes centrifuged in this study. Solutes are ordered by the molar concentration required to transition 50% of the solution in Fig. 2b.** Solutes at the bottom of the table affected the observed centrifugal mobility of all solution solutes more sensitively per mole than those at the top of the table.

| Solute | $M_s$, g/mol | $\bar{v}_s$, cm³/mol* | $z_s$, charge | $\frac{\partial \rho_{soln}}{\partial c_s}$, g/mol | $c_t$, for 50% transition (mM) at 60 kRPM |
|---|---|---|---|---|---|
| H$_2$$^{18}$O | 20.0 | 0.0 | 0 | 2.0 | 6000 ± 1500 |
| TEGDME | 222.3 | 206.8 | 0 | 15.5 | 700 ± 300 |
| MeCN (Acetonitrile) | 41.1 | 47.4 | 0 | 6.3 | 380 ± 30 |
| EtOH | 46.1 | 55.1 | 0 | -9.0 | 350 ± 50 |
| Li$^+$Cl$^-$ | 6.9, 35.5 | -4.7, 21.6 | +1, -1 | 11.6, 13.9 | 130 ± 20 |
| Na$^+$OH$^-$ | 23.0, 17.0 | -5.0, -0.2 | +1, -1 | 28.0, 17.2 | 28 ± 5 |
| Li$^+$I$^-$ | 6.9, 126.9 | -4.7, 40.0 | +1, -1 | 11.6, 86.9 | 5 ± 1 |
| CH$_2$Br$_2$ | 173.8 | 68.4 | 0 | 105.4 | 4.8 ± 1.0 |
| Mg$^{2+}$SO$_4$$^{2-}$ | 24.3, 96.1 | -28.8, 21.6 | +2, -2 | 53.1, 74.5 | 3.0 ± 0.8 |
| K$^+$I$^-$ | 39.1, 126.9 | 5.2, 40 | +1, -1 | 33.9, 86.9 | 2.5 ± 0.5 |
| K$_2$$^+$SO$_4$$^{2-}$ | 39.1$_2$, 96.1 | 5.2$_2$, 21.6 | +1, -2 | 33.9$_2$, 74.5 | 2.0 ± 0.5 |
| Cs$^+$I$^-$ | 132.9, 126.9 | 17.5, 40.0 | +1, -1 | 115.4, 86.9 | 0.9 ± 0.2 |

*(25°C values, $\frac{\Delta \bar{v}_s}{\Delta T} \approx 5 \times 10^{-4} cm/g/K$)

** The two numbers in $M_s$, $\bar{v}_s$ and $\frac{\partial \rho_{soln}}{\partial c_s}$ represent cation and anion in sequence. $\bar{v}_s$ are obtained from "Durchschlag, H. & Zipper, P. Calculation of the partial volume of organic compounds and polymers. *Progress in Colloid & Polymer Science* 94, 20-39 (1994)"[3]. The equation $\frac{\partial \rho_{soln}}{\partial c_s} = M_s(1 - \bar{v}_s \rho_{solvent})$ is then used for the partial molar density.

## Section S4. Centrifugation Simulation Methods

**4.1. Solving Equations:** The governing flux equations, given in Eq. 1, Eq. 2, and Eq. 3 of the manuscript were solved numerically in MATLAB. The centrifuge can be discretized radially in the $r$ dimension, representing the radius. The centrifuge is symmetric in all other dimensions. The $r$-direction is discretized from the inner to outer radii, $[r_{inner}, r_{outer}]$, in $m_0$ intervals. For each interval, the concentration is defined at the center and the flux is defined on the boundary so that mass is conserved.

The flux at each location is calculated by using a discretized Eq. 1, and then the concentration for the next timestep is calculated by using a discretized Eq. 2 (forward Euler). This process is then repeated for many timesteps until the rate of change of the system is sufficiently small and it has reached equilibrium. Each term of Eq. 1 can be discretised separately, for example, the first diffusion term is calculated as Eq. S1.

$$J_{F,r}(m) = D_i \frac{c_i(m) - c_i(m-1)}{dr} \quad m = 2, \ldots, m_0 \quad (S1)$$

Activity coefficient effects were neglected for all solutions in the simulations. Electrostatic fluxes, where relevant, are calculated by imposing charge neutrality on the solution. More details of this method can be found in the Supporting Information of *Wild, et al. (2023)*[4]. The two boundaries of the centrifuge are imposed by defining the fluxes into or out of the centrifuge tubes as zero. For example, at the inner radii, $J_r(1) = 0$. Once each flux term is calculated at each location, Eq. 2 in its radial form, Eq. S2, is discretized and used to calculate the next array of concentrations, Eq. S3.

$$\frac{\partial c_i}{\partial t} = -\nabla \cdot \vec{J}_i = -\frac{\partial (rJ_{i,r})}{\partial r} \quad (S2)$$

$$c_i(m, t+dt) = c_i(m,t) - dt \left[ \frac{r(m)J_{i,r}(m) - r(m-1)J_{i,r}(m-1)}{2r_c(m)dr} \right] \quad (S3)$$

**4.2. Model Inputs, Outputs, and Assumptions:** A summary of the model inputs and assumptions are given in Table S3. All arrays and variables are initialized at the outset for computational efficiency. The primary output is the final concentration distribution arrays, $c_i(r)$, at the final time, which define the separation factor at all locations.

**Table S3. Model inputs and assumptions.**

| Input | Definition and Dimension | Assumptions |
|---|---|---|
| $\omega$ | Angular velocity (scalar) | Constant – no ramping |
| $\rho_i$ | Density (scalar) | Solutions are incompressible. Isotope species density is proportional to $M_i$ |
| $dr, dt,$ | Discretized dimensions (scalar) | Uniform spacing |
| $c_i(r,t)$ | Molar concentration (matrix) | Uniform at $t = 0$ for each species |
| $D_i$ | Diffusion coefficient (scalar) | Constant – no pressure or other dependencies |
| $J_i(r,t)$ | Molar flux (matrix) | No fluxes in to or out of boundaries (see below) |
| $M_i$ | Molecular mass (scalar) | - |
| $r$ ($r_{inner}$ and $r_{outer}$) | Centrifuge radii (scalar) | - |
| $T$ | Temperature (scalar) | Constant and uniform |
| $\bar{v}$ | Partial specific volume (scalar) | Constant – no pressure or other dependencies |

**4.3. Model Boundary Conditions and Initial Conditions:** The initial condition is a uniform distribution of all solute species in the centrifuge, while the boundary conditions are no fluxes through any boundary. For example, there are no radial fluxes at the inner and outer radius boundary. All conditions are expressed below.

$$c_i(1:m_0, t = 1) = c_{i,0}$$

$$J_{r,i}(1) = J_{r,i}(m_0) = 0$$

## Section S5. Centrifugation Simulation Results and Prior Experiments

Centrifugal simulations using the methods of Section S4 show that the natural logarithm of the transient separation factor is linearly proportional to the ratio of $u_s/D_s$ to high precision, and therefore the transient state can equally be used to determine the ratio (Fig. S2).

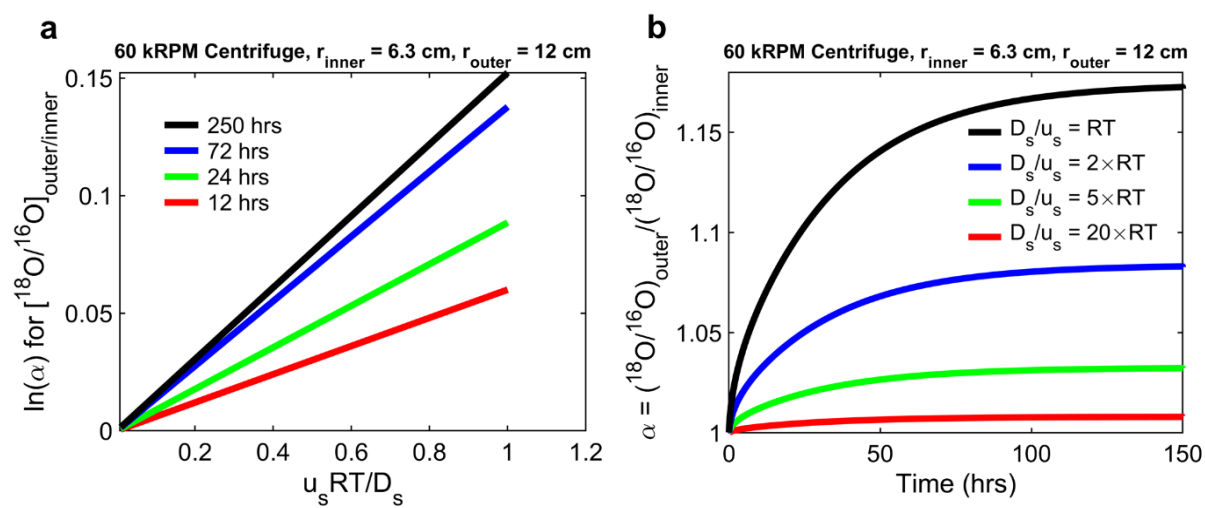

**Figure S2** – (a) The separation factor α of $^{18}O/^{16}O$ as a function of $u_s RT/D_s$ after 12 hrs, 24 hrs, 72 hrs, and 250 hrs (~equilibrium), showing a linear relation between $ln(\alpha)$ and $u_s RT/D_s$. A is defined as $\alpha_o = (c_{18_O}/c_{16_O})_{outer\ radius}/(c_{18_O}/c_{16_O})_{inner\ radius}$. (b) The transient separation factor of $^{18}O/^{16}O$ for $u_s RT/D_s$ = 0.05, 0.2, 0.5, and 1.0. The centrifuge conditions are given in the figure title and the water properties at 40°C are used.

Table S4, reproduced from the Supporting Information of *Wild, et al. (2023)*[4], shows good experimental agreement with the case of $u_s RT/D_s = 1$, which aligns with the observation of this study since a salt concentration of 0.5 mol/kg was used there.

**Table S4** - $^{16}O/^{18}O$ Separation factors

| Sample | Time (Hours) | Inner Separation Factor | Outer Separation Factor | Total Separation Factor | Theoretical α if $u_s RT/D_s = 1$ |
|---|---|---|---|---|---|
| Water (0.5 m LiCl), 1 | 24 | 1.0388 | 0.9552 | 1.0875 | 1.0903 |
| Water (0.5 m LiCl), 2 | 24 | 1.0382 | 0.9541 | 1.0882 | 1.0903 |
| Water (0.5 m LiCl), 3 | 24 | 1.0380 | 0.9539 | 1.0881 | 1.0903 |
| Water (0.5 m LiCl), 1 | 72 | 1.0586 | 0.9331 | 1.1342 | 1.1426 |
| Water (0.5 m LiCl), 2 | 72 | 1.0588 | 0.9341 | 1.1335 | 1.1426 |

**Section S6. Inner and Outer Separation Factor**

In addition to the overall separation factor of $^{18}O/^{16}O$ measured in the centrifuge tube as shown in Fig. 2a of the main text, the individual separation factor at the inner and outer radii are given in Figure S3. The inner separation factor is defined as $([^{18}O]/[^{16}O])_{inner}/([^{18}O]/[^{16}O])_{natural}$ which corresponds to the centrifuge inner radius, and the outer separation factor is defined as $([^{18}O]/[^{16}O])_{outer}/([^{18}O]/[^{16}O])_{natural}$, which corresponds to the centrifuge outer radius. $([^{18}O]/[^{16}O])_{natural}$ means $[^{18}O]/[^{16}O]$ in a solution before centrifugation. The inner and outer separation factors are approximately symmetric.

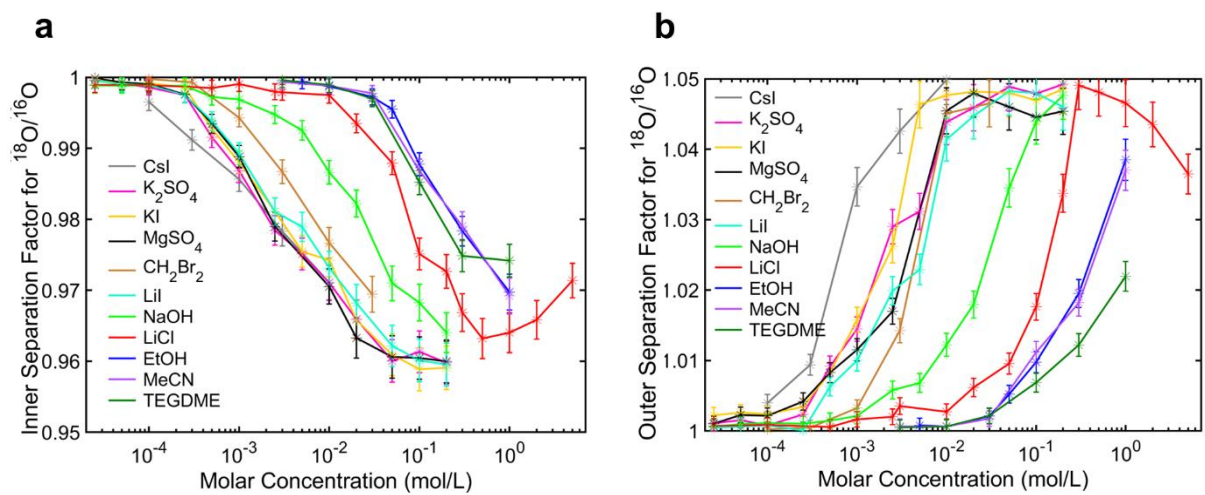

**Figure S3** – (a) The inner separation factor of $^{18}O/^{16}O$ with different solutes. (b) The outer separation factor of $^{18}O/^{16}O$ with different solutes. The centrifugal speed is 60 kRPM and the temperature is 40°C.

**Section S7. Salt Concentration Polarization**

The concentration polarization of dissolved salts after centrifuging was determined by measuring the conductivity of the same samples at the inner and outer radii of the centrifuge as were used to measure the $^{18}O/^{16}O$ separation. A LAQUAtwin Compact Water Quality Conductivity Meter was used. Before each set of measurements, the meter was calibrated using the supplied 1.41 mS/cm reference. The stated precision of the Meter is '±2% full scale. ±1 digit (for each range): ±5 µS/cm (0 to 199 µS/cm) - ±0.05 mS/cm (0.20 to 1.99 mS/cm)'. In general, 100 µL of the sample was placed over the meter sensor for measurement. The sample was left for 10-15 seconds to obtain a stable reading. The sample would then be discarded before cleaning the sensor with deionized water and drying the sensor. The precision of the conductivity meter was compared to an inductively coupled plasma mass spectrometry (ICP-MS) for precision evaluation and showed that the conductivity meter gives good linearity to concentration between 0.1 mM and 50 mM.

The separation factor of the dissolved solutes was measured by comparing the conductivity of the outer radius sample to the inner radius sample. For example, a separation factor value of 2.0 means that the salt concentration at the outer radii was 2.0x the concentration at the inner radii after centrifugation. Figure S4 shows the salt separation factor results for salts at various centrifugal speeds after 24 hrs.

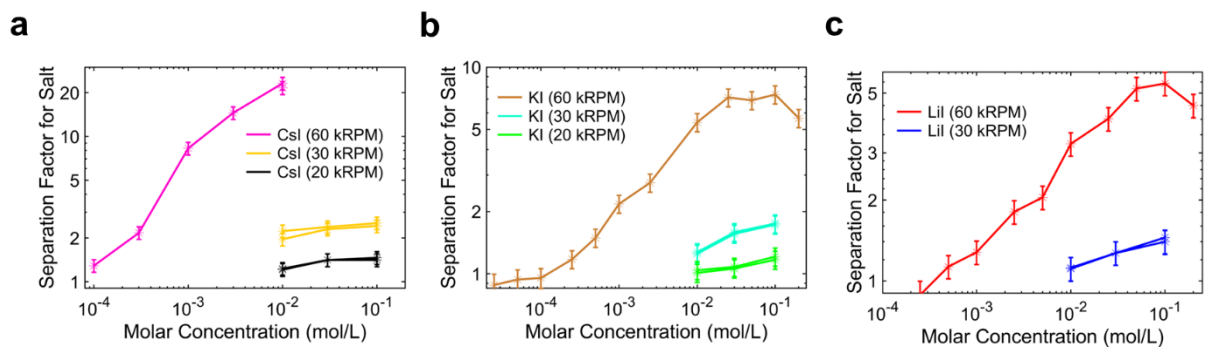

**Figure S4** - Experimental results for the solute polarization at different centrifugal speeds and salt concentrations for dense solutes. (a) CsI, (b) KI, and (c) LiI.

Figure S5 shows the non-normalized salt separation factors from Fig. 2D in the main text.

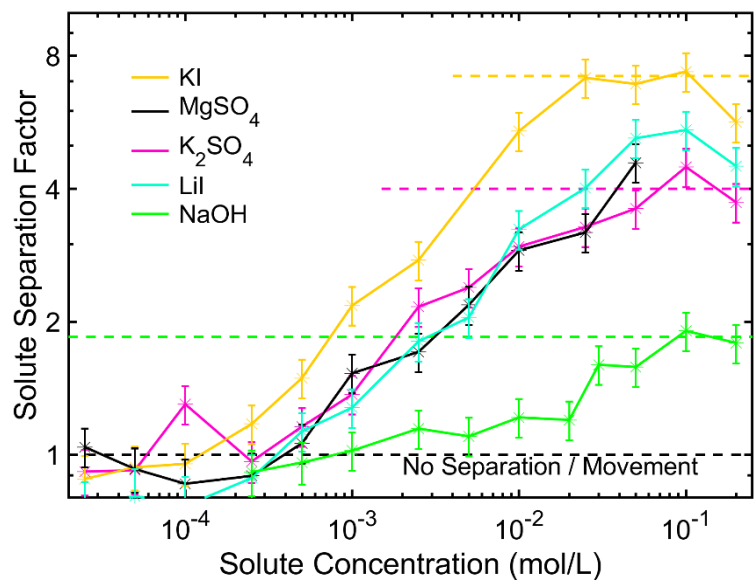

**Figure S5** - The dependence of solute separation factor on solute concentration in the aqueous solution. This factor is defined as the ratio of solute concentration at the outer and inner radii of a centrifuge tube. The dash lines indicate the theoretical separation factor if the ER is valid.

## Section S8. Effects of Centrifugal Field Strength

Several dense salts (CsI, KI, and LiI) were centrifuged for 24 hours at varying speeds to investigate if the centrifugal acceleration (g) influenced the critical concentration for the transition. Both water isotope separation and salt polarization were measured. The experimental results are shown in Figures S6 and S7, respectively.

We found that if the centrifugal acceleration effect is not considered, the model ($V_{0,i} = A_0 \left(\frac{\partial \rho_{soln}}{\partial c_i}\right)^2$) does not fit all results at different g well (Fig. S6b,d,f). However, once the effect is considered the model can fit centrifugal results at different rotation speed very well. Such results suggest that internal stress in water may contribute to the observed ER breakdown.

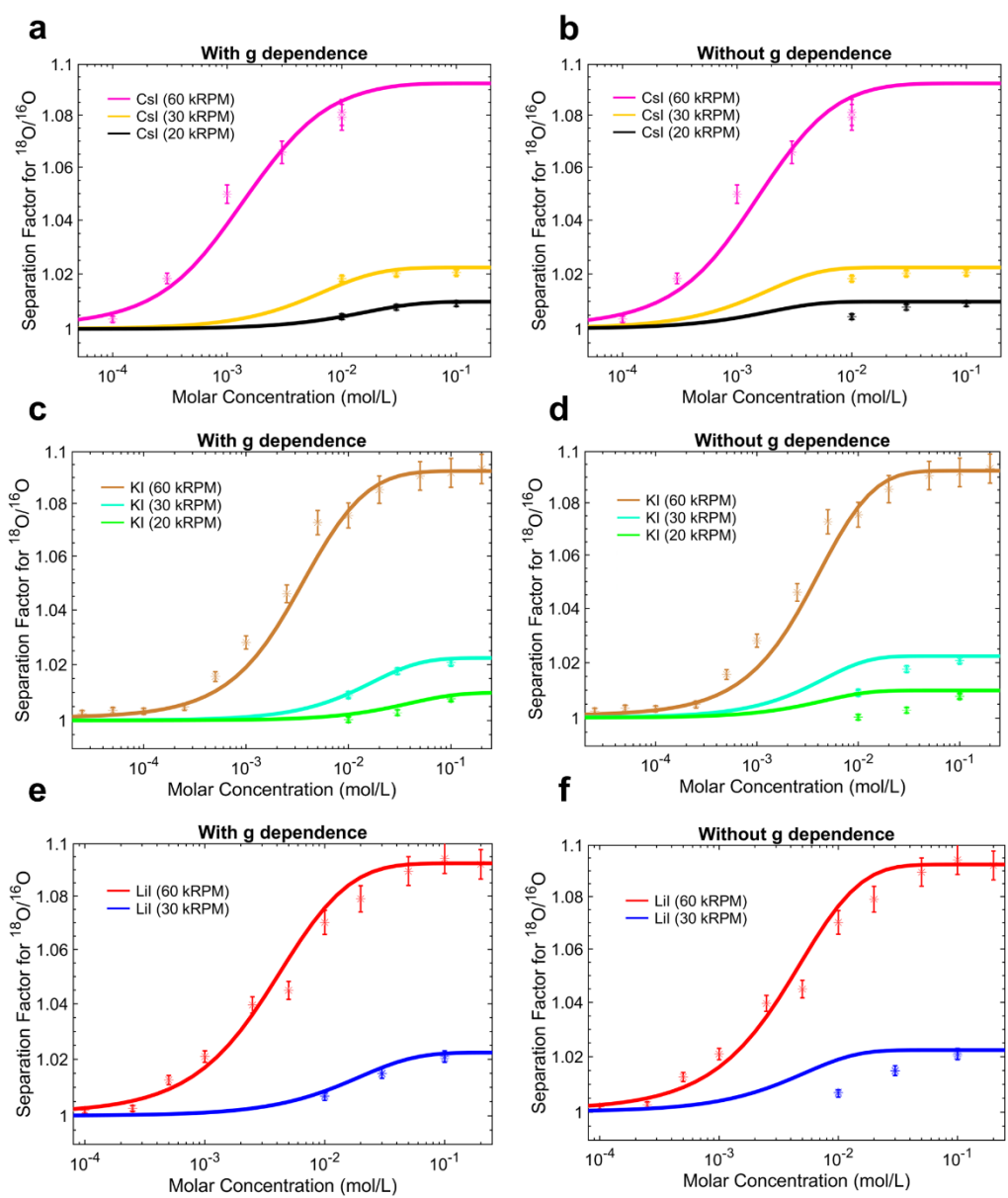

**Figure S6** - Experimental results for the **water isotope separation** at different centrifugal speeds. The solid lines are the fitting curves with centrifugal acceleration (*g*) dependence as in Eq. 4 in the main text (a,c,e) and without the dependence (b,d,f).

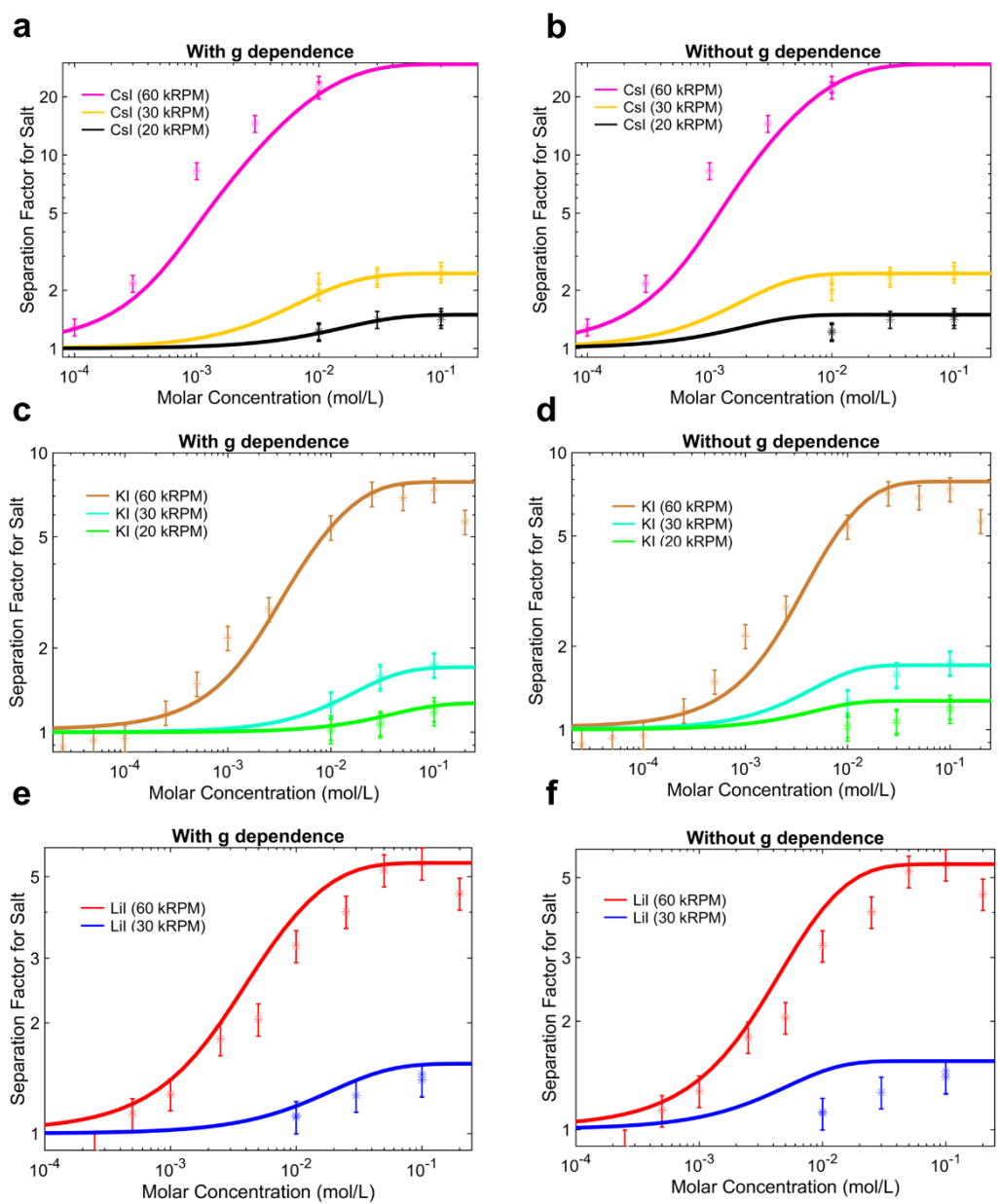

**Figure S7** - Experimental results for the **salt concentration polarization** at different centrifugal speeds. The solid lines are the fitting curves with centrifugal acceleration dependence (*g*) as in Eq. 4 in the main text (a,c,e) and without the dependence (b,d,f).

## Section S9. JMAK Analysis / Avrami Equation Analogy

Johnson-Mehl-Avrami-Kolmogorov (JMAK) theory is traditionally used to calculate the volume fraction which has undergone crystallization or phase change during some non-equilibrium process. Typically, this can describe the sigmoid-shape of volume-transformed as a function of time, for example as shown in Figure S8. Here, its ideas are used to describe the volume transformed as a function of concentration, where the transformed volume surrounding each solute is fixed.

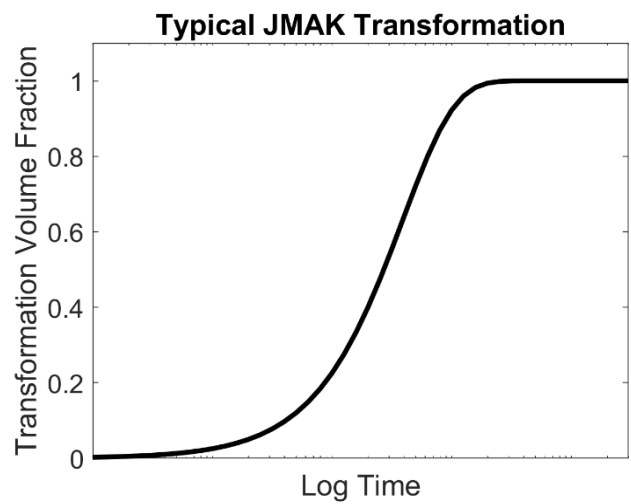

**Figure S8** - A characteristic time-dependent transformation of material under crystallized described by JMAK theory.

Within JMAK theory, there are two 'volume fractions' considered: The first is $f_{transformed}$ which represents the actual volume fraction transformed and therefore takes a value $0 \leq f_{transformed} \leq 1$ ($V_1/V_{total}$ in main text), and then there is $f_{extended}$, which blindly sums up the volume of all transition spheres regardless of if they are overlapping, and divides this volume by the region $V_{total}$, resulting in $0 \leq f_{transformed} \leq \infty$ ($\sum_i V_{0,i} c_i / V_{total}$ in main text). A simple illustration of the two are given in Figure S9.

The relation $f_{transf} = 1 - exp(-f_{extended})$ comes from solving the simple expression $df_{transformed} = (1 - f_{transformed}) df_{extended}$. Here, $df_{transformed}$ is the incremental change in the transformed volume fraction, $1 - f_{transformed}$ is the fraction of space that is untransformed, and $df_{extended}$ is the incremental extended volume fraction.

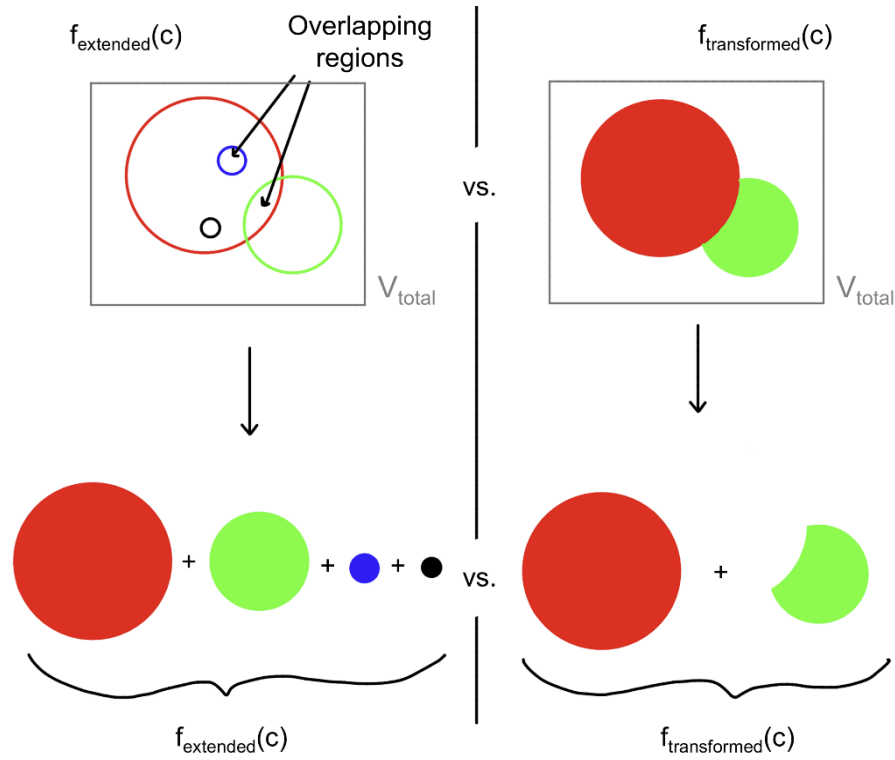

**Figure S9** - Illustration in 2D of the two volume fractions considered in JMAK theory. $V_{total}$ is the total area within the boundary box, while the two volume fractions are the summed areas of the highlighted regions relative to this. $f_{extended}$, on the left, sums of volume of all enclosed regions, regardless of whether they are overlapping with an existing region, meaning that some areas count multiple times. While $f_{transformed}$, on the right, sums strictly the regions transformed at all, and overlapping regions do not count more than once.

As described in the manuscript, 'the effective centrifugal mobility of a species, s, in the whole solution is the volume-weighted average of these two phases.' This follows from that the solute separation results showed that dense solutes, such as CsI, would not themselves drift in very dilute solutions. Therefore, the transition sphere surrounding a solute itself cannot alone allow the solute to move freely. Instead, the entire local region would appear to matter, which may include millions of water molecules. The effect of this on the flux equation is given in Eq. S4.

$$\vec{J}_s = -D_s \nabla c_s + f_{transformed} D_s \frac{\omega^2 \vec{r}}{RT} c_s \frac{\partial \rho_{soln}}{\partial c_s} \quad (S4)$$

As shown in Fig. 3b of the manuscript, this model quantitatively matches the experimental shape of the transition curve, thereby indicating that the model assumptions are met. Figure S10 shows the inner and outer radii results, which also show that the model prediction matches experimental results well. Figure S11 shows the model results for other solutes tested in this paper.

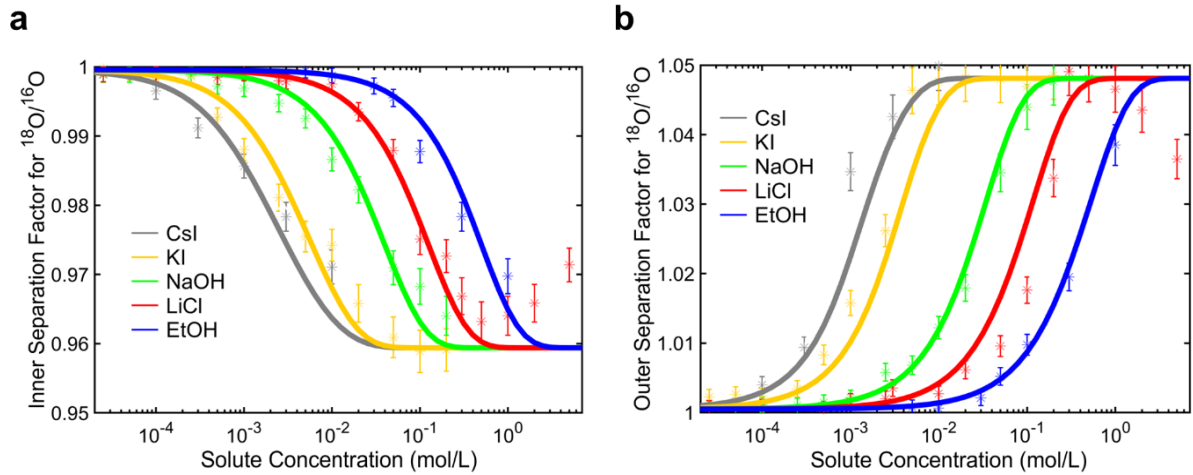

**Figure S10** - The simulated curves (thick lines) overlayed with the inner (a) and outer (b) radii experimental results of using the best fit correlation to determine the transition sphere volume. The same $A_0$ of $5.57 \times 10^{-6}$ m$^2 \cdot$s$^2 \cdot$mol/kg$^2$ is used in fitting all curves.

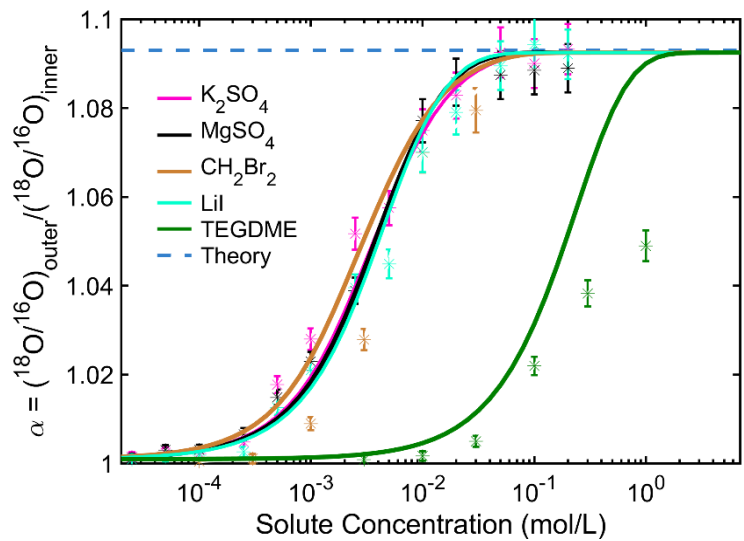

**Figure S11** - Experimental results of the concentration-dependent α for remaining solutes in Fig. 2D, together with fitting curves from the single-fitting-parameter model described by Eq. 6.

## Section S10. Results at a Low Centrifugal Speed

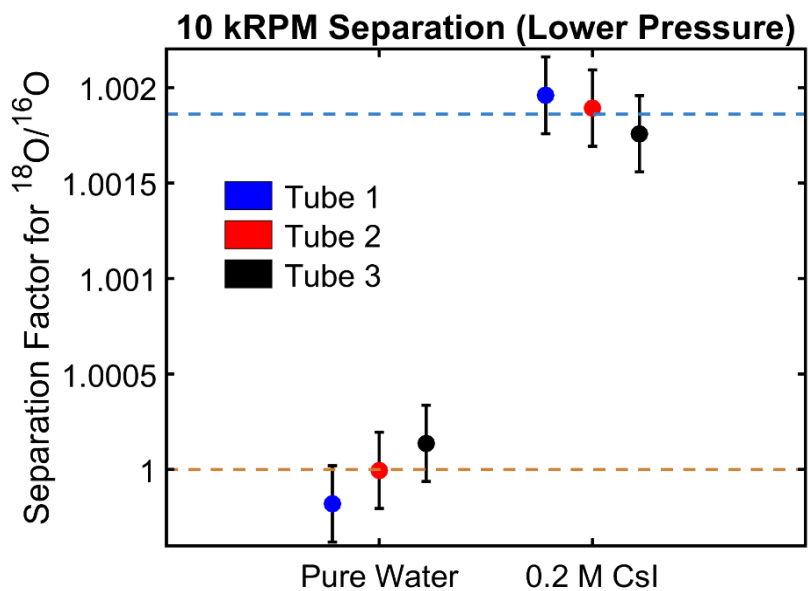

**Figure S12** – $^{18}O/^{16}O$ isotope separation factor for both pure water and 0.2 M CsI after 24 hrs at 10 kRPM, where the maximum pressure inside is only 6 MPa. A large reduction is the separation factor is again seen. This indicates that pressure is unlikely to cause the phenomenon. The blue line represents theoretical value based on Eq.1 and 2 assuming the ER holds.

## Supporting References